# THE FIRST TURBULENT COMBUSTION

CARL H. GIBSON[1]
Departments of Mechanical and Aerospace Engineering and Scripps Institution of Oceanography, University of California, San Diego
La Jolla, California 92093-0411, USA

The first turbulent combustion arises in a hot big bang cosmological model Gibson (2004) where nonlinear exothermic turbulence permitted by quantum mechanics, general relativity, multidimensional superstring theory, and fluid mechanics cascades from Planck to strong force freeze out scales with gravity balancing turbulent inertial-vortex forces. Interactions between Planck scale spinning and non-spinning black holes produce high Reynolds number turbulence and temperature mixing with huge Reynolds stresses driving the rapid inflation of space. Kolmogorovian turbulent temperature patterns are fossilized as strong-force exponential inflation stretches them beyond the scale of causal connection *ct* where *c* is light speed and *t* is time. Fossil temperature turbulence patterns seed nucleosynthesis, and then hydro-gravitational structure formation in the plasma epoch, Gibson (1996, 2000). Evidence about formation mechanisms is preserved by cosmic microwave background temperature anisotropies. CMB spectra indicate hydro-gravitational fragmentation at supercluster to galaxy masses in the primordial plasma with space stretched by $\sim 10^{50}$. Bershadskii and Sreenivasan (2002, 2003) CMB multi-scaling coefficients support a strong turbulence origin for the anisotropies prior to the plasma epoch.

## INTRODUCTION

The current cosmological model requires a hot, explosive, initial-singularity of the universe called the Big Bang. Sir Fred Hoyle invented the Big-Bang terminology in a

[1] cgibson@ucsd.edu, http://www-acs.ucsd.edu/~ir118

1950 BBC radio show, not to propose a first-turbulent-combustion beginning of the universe but to ridicule the idea. Hoyle's difficulty with the big bang concept had partly to do with its orthogonality to his own cosmological theory that the mass-energy of the universe should be created continuously (Hoyle, 1994; Hoyle, Burbidge and Narlikar, 2000). Hoyle also felt the extreme remoteness in space and time and the unknown physics of this bizarre singularity would forever prevent adequate observational testing. Quoting Dirac, Hoyle (1994) disparages any big bang beginning of the universe by suggesting "that which cannot be observed does not exist".

New observations, new physics, and new fluid mechanics may have resolved Hoyle's presumption that a big bang origin for the universe is observationally untestable. Although perhaps we can't observe an actual big bang, just as we presumably can't observe living dinosaurs, we can believe that both dinosaurs and the big bang existed if we can find enough of their fossils and develop convincing paleontologies. It is now the epoch of precision cosmology, with information pouring in from numerous ground and space based telescopes. Hoyle and colleagues Bondi and Gold based their 1948 continuous-creation-cosmology on a scalar C term added to the Einstein general relativity (GR) equations, just as Einstein once added a cosmological constant $\Lambda$ (in what he later decided was his "biggest blunder") to explain an apparent lack of expansion of the universe. Cosmologists still tinker with the GR equations to explain apparent expansion accelerations by "dark energy", "Cardassian cosmologies", "quintessence" and "cosmic jerks" (Riess et al., 2004) where the universe decelerates its expansion for 7 billion years and then accelerates. These adjustments to GR and cosmology will likely vanish with $\Lambda$ as new data arrives and as new physics and new fluid mechanics (hydro-gravitational-dynamics, or HGD) are included in the analysis (Gibson, 1996). The large population of gassy "grey dust" planetary objects predicted by HGD as the baryonic dark matter can easily account for the slight dimming of supernovae Ia that lead to all these adjusted GR models. Because dark planets accrete in pairs to produce pairs of small stars, when one binary dies its companion supplies the gas that grows the white dwarf ashes to a dense explosive mass of spinning carbon. Plasma jets evaporate a halo of planet atmospheres to dim the eventual supernova on some lines of sight (Gibson and Schild, 2004).

Another reason some might question a strong-turbulent beginning at small big-bang scales is that the "standard-model" of turbulence (in which turbulence must cascade from large scales to small) rules this out. Evidence of strong big-bang-turbulence emerges from recent observations, but not evidence of strong "standard-model" turbulence. Thus, the standard model of turbulence must be revised to permit big-bang-turbulence (Gibson, 2000). In the laboratory, ocean, and atmosphere, turbulence always starts at viscous-inertial-vortex Kolmogorov scales and cascades to larger scales, dominated and controlled at all stages by inertial-vortex-forces $\vec{v} \times \vec{\omega}$, where $\vec{v}$ is the velocity and $\vec{\omega}$ is the vorticity. Big bang turbulent combustion begins at the smallest scale known to physics, the Planck scale, and fossils of this seminal turbulence have now expanded to scales larger by a factor of $10^{85}$.

This necessary re-definition of turbulence explains the well-documented universal scaling laws of Kolmogorov, Batchelor, Obukhov and Corrsin (Gibson, 1991). Moreover, the intrinsic irreversibility and entropy production of turbulence (as redefined based on $\vec{v} \times \vec{\omega}$) results in persistent fingerprints in a variety of hydrophysical fields, termed *fossil turbulence* (Gibson, 1999). Fossil-turbulence remnants preserve information about previous turbulence events for much longer time periods than the turbulence (as redefined). Cosmic Microwave Background (CMB) temperature anisotropies $\delta T$ represent the most ancient fossils of the primordial universe. Small CMB $\delta T/T \sim 10^{-5}$ fluctuations have been subjected to intense scrutiny by a host of microwave-telescopes on the ground, on balloons, and in space. The Wilkenson Microwave Anisotropy Probe (WMAP) telescope has been orbiting the second Lagrange point of the earth-sun system more than a year collecting data. In our fossil big-bang-turbulence-combustion model, the life of the initial big-bang-turbulence event is only $10^{-35}$ s, or $10^{8}$ Planck time periods. For a persistence time of $\sim 4.3 \times 10^{17}$ s (>13 Gyr), we find a big-bang-turbulence fossil/event duration ratio of more than $10^{52}$.

Clear evidence of strong primordial turbulence emerges from the CMB $\delta T$ maps (Bershadskii and Sreenivasan 2002, 2003). A hot big-bang-turbulence (BBT) model has been presented (Gibson, 2004) based on these and other data and recent cosmological physics (Peacock, 2000). Physical parameters of the improbable, exothermic, strong-turbulence big-bang event are fossilized by GR inflation of space-time stretching the resulting temperature fluctuations beyond the scale *ct* of causal connection, where *c* is the speed of light and *t* is time. Hot BBT is quite efficient. Stars burn <1% of their hydrogen mass to form helium, but Planck-scale turbulent combustion burns ~42% of the particle rest mass-energy to produce a renewed gas of Planck-particle fuel as its ashes.

In the following paper, we discuss turbulence and fossil turbulence definitions and the turbulence cascade direction. Then we review the physical processes of cosmology and particle physics that suggest a big-bang-turbulence-combustion physical mechanism. Dimensional analysis is applied to the quantum-gravitational-dynamics (QGD) epoch at Planck scales since a small number of relevant dimensional parameters exist and the process is intrinsically nonlinear. In the conventional approach, "natural units" obtained by setting the relevant parameters *c*, *h, G* and *k* to 1 are used to simplify linearized equations, where *c* is light speed, *h* is Planck's constant, *G* is Newton's constant, and *k* is Boltzmann's constant (Weinberg, 1972). To understand the CMB evidence of big-bang-turbulence it is necessary to abandon natural units and the simplifying assumptions of linearity and collisionless-fluid-mechanics used in standard-model cosmology. In particular, using "cold-dark-matter" (CDMHC) hierarchical clustering of CDM "seeds" or "halos" to explain gravitational structure formation is unacceptable, along with the Jeans 1902 acoustical criterion for gravitational instability.

## SAMPLING HYDRO-GRAVITATIONAL AND TURBULENCE FOSSILS

Evidence and equations leading to an inertial-vortex force based definition of turbulence and fossil turbulence to describe big bang turbulence and the initial stages of cosmology are provided in a related paper (Gibson, 2004). Universal similarity theories of turbulence and turbulent mixing have strong experimental support for second order statistical parameters like power spectra for laboratory flows, direct numerical simulations, and natural turbulent flows in the atmosphere and ocean (Gibson, 1986) as well as in the Galaxy (Gibson, 1991). Deviations from precise universal turbulence similarity arise from increasing intermittency of dissipation rates with increasing

Reynolds number. Self-gravitational structure formation is also a nonlinear process starting at small scales, also producing an extremely intermittent distribution (density rather than vorticity) difficult to sample at small scales without error.

Failure to recognize the large undersampling errors intrinsic to intermittent high Reynolds turbulence and a failure to take advantage of information preserved by fossils of turbulence have resulted in the dark mixing paradox of ocean microstructure studies (Leung and Gibson, 2004). Similarly, planetary mass baryonic (ordinary) dark matter objects have eluded intensive star microlensing searches by three collaborations searching for massive galaxy halo objects (MACHOs) because all three assumed a uniform MACHO density distribution rather than the highly clumped and patchy distribution expected from HGD (Gibson and Schild, 1999). HGD predicts that after a period of gravitational fragmentation from large scales to small during the hot plasma epoch before transition to gas, the plasma turns into a primordial fog of particles in trillion particle clumps that persist as the baryonic dark matter (Gibson, 1996). The nonlinear self-gravitational cascade of these planetary mass particles to stellar mass produces an intermittent lognormal distribution of the particle density that is difficult to sample directly. Quasar microlensing by a foreground galaxy show from the twinkling frequency of the quasar images that the dominant mass component of the lens galaxy must be planetary mass objects (Schild, 1996).

Thirty years of microstructure sampling in the deep ocean have failed to sample any of the dominant turbulent microstructure patches that produce the main thermocline, although a few of their fossils have been detected (Gibson, 1999). Recently it has been claimed that turbulence mixing and diffusion in equatorial waters approach a minimum rather than a maximum (Gregg et al., 2003) despite the fact that maximum biological productivity at low latitudes requires a maximum in the turbulent mixing and diffusion at the equator, not a minimum. Rare, large-vertical-scale fossils of turbulence observed at depths with strong stratification confute claims that equatorial turbulence is minimal. Microstructure undersampling errors increase to a large maximum at the equator because Coriolis forces vanish that otherwise limit horizontal turbulence length scales (Baker and Gibson, 1987). Likewise, fossils of turbulence and early gravitational structure formation detected in the temperature anisotropies of the cosmic microwave background preserve information about the hydro-gravitational evolution of the universe after its big bang turbulence combustion origin.

## THE FIRST TURBULENT COMBUSTION

In our turbulent combustion model of the hot-big-bang universe (Gibson, 2004), space-time-energy-entropy emerges from the vacuum spontaneously and explosively at Planck scales. Small scales are described by quantum mechanics (QM) and large scales by general relativity (GR) theories of physics. Both QM and GR theories begin to fail at Planck scales. Multi-dimensional super-string (MS) theory shows promise of QM and GR reconciliation (Greene, 1999). Additional Planck scale insights provided by universal similarity laws of turbulence and turbulent mixing, thermodynamics and fluid mechanics (FM) are not considered by GR, QM or MS. FM is unique among these physical theories because it applies over the full range of scales of the universe and cannot be ignored at any scale.

The quantum gravitational dynamics (QGD) epoch starting at the Planck time must be described by QM, GR, MS and FM theories using appropriate analytical methods for this process where some non-linear mechanism dominates that must supply efficient entropy and energy production, and where a small number of dimensional parameters are relevant. Dimensional analysis that applies the parameters to a physical model based on available theory is required, as in Figures 1 and 2. The initial quantum-tunneling Planck event is allowed by Heisenberg's uncertainty principle of QM, where the uncertainty of the energy of a particle $\Delta E$ multiplied by the uncertainty of its time of existence $\Delta t$ is greater than or equal a constant $h$ termed the Planck constant, Fig. 1. The Planck mass $m_P = (ch/G)^{1/2}$ is found by equating the QM Compton wavelength $L_C = h/mc$ of a particle with mass $m$ to the GR Schwarzschild radius $L_S = Gm/c^2$ of a black hole with the same mass, where $c$ is the speed of light and $G$ is Newton's gravitational constant. The Planck entropy $S_P$ is equal to the Boltzmann constant $k$. This gives a minimum black hole specific entropy $s_P = S_P/m_P$, maximum black hole temperature $T_P = (c^5h/Gk^2)^{1/2}$, and minimum black hole evaporation time $t_P = (c^{-5}hG)^{1/2}$ (Greene, 1999). Grand unified (GUT) force theories suggest all basic forces of nature (weak, strong, electromagnetic, gravity, inertial-vortex) are equivalent at the Planck-GUT temperatures ($10^{32}$ to $10^{28}$ K) and are quantized by vibrations of multidimensional string-like-objects at Planck scales from MS. The string tension is the Planck force $F_P = c^4G^{-1}$ in Fig. 1. This force group appears in Einstein's GR equations (3) to normalize the stress-energy tensor.

**Heisenberg uncertainty:** $[\Delta E \times \Delta t]_P = E_P t_P \geq h$
**Compton wavelength:** $L_{Compton} = L_C = h/mc$
**Schwarzschild radius:** $L_{Schwarzschild} = L_S = Gm/c^2$
*Note:* $L_C = L_S = L_P\ @\ m = m_P$

**Planck scale physical constants:**
$c=2.998\times10^{8}$ m s$^{-1}$ ; $h=1.05\times10^{-34}$ kg m$^{2}$ s$^{-1}$ ;
$G=6.67\times10^{-11}$ m$^{3}$ kg$^{-1}$ s$^{-2}$ ; $k=1.38\times10^{-23}$ kg m$^{2}$ s$^{-2}$ K$^{-1}$

**Planck scales:**

| Mass | Length | Time | Temperature |
|---|---|---|---|
| $m_P = [chG^{-1}]^{1/2}$ | $L_P = [c^{-3}hG]^{1/2}$ | $t_P = [c^{-5}hG]^{1/2}$ | $T_P = [c^{5}hG^{-1}k^{-2}]^{1/2}$ |
| $2.12\times10^{-8}$ kg | $1.62\times10^{-35}$ m | $5.41\times10^{-44}$ s | $1.40\times10^{32}$ K |

**Planck big-bang-turbulence scales:**

| Energy | Power | Dissip. | Entropy | Density | Force | Grav.,Turb. | Stress |
|---|---|---|---|---|---|---|---|
| $E_P$ | $P_P$ | $\varepsilon_P$ | $s_P$ | $\rho_P$ | $F_P$ | $g_P$, $[\vec{v}\times\vec{\omega}]_P$ | $\tau_P$ |
| $[c^{5}hG^{-1}]^{1/2}$ | $c^{5}G^{-1}$ | $[c^{9}h^{-1}G^{-1}]^{1/2}$ | $[c^{-1}h^{-1}k^{2}G]^{1/2}$ | $c^{5}h^{-1}G^{-1}$ | $c^{4}G^{-1}$ | $[c^{7}h^{-1}G^{-1}]^{1/2}$ | $[c^{13}h^{-3}G^{-3}]^{1/2}$ |
| $1.94\times10^{9}$ | $3.64\times10^{52}$ | $1.72\times10^{60}$ | $6.35\times10^{-16}$ | $6.4\times10^{96}$ | $1.1\times10^{44}$ | $5.7\times10^{51}$ | $2.1\times10^{121}$ |
| kg m$^{2}$ s$^{-2}$ | kg m$^{2}$ s$^{-3}$ | m$^{2}$ s$^{-3}$ | m$^{2}$ s$^{-2}$ K$^{-1}$ | kg m$^{-3}$ | kg m s$^{-2}$ | m s$^{-2}$ | m$^{-1}$ s$^{-2}$ |

Figure 1. Planck scales and Planck parameters of big-bang-turbulence-combustion.

Planck mass, length, time, and temperature scales are found by dimensional analysis from the fundamental constants *c, h, G,* and *k*. Planck energy, power, viscous dissipation rate, specific entropy, density, stress and gravitational force scales of Fig. 1 all give appropriate physical values, so the model in Fig. 2 is supported. For example, the large Planck inertial-vortex-force per unit mass of big-bang turbulence $[\vec{v}\times\vec{\omega}]_P = L_P/{t_P}^2$ equals the Planck gravitational force per unit mass $g_P = F_P/m_P = \left[c^7/hG\right]^{1/2} = 5.7\times10^{51}$ m $s^{-2}$ (Fig. 1). The turbulent Reynolds stress $2.1\times10^{121}$ $m^{-1}$ $s^{-2}$ supplies so much negative pressure that turbulence emerges as a candidate to drive exponential space-time inflation following the Einstein GR equations, along with gluon viscous stresses and the false vacuum mechanism (Guth, 1997). The Planck power $P_P$ exceeds that of all the stars in the present horizon scale *ct*.

Figure 2 suggests big bang turbulent combustion (Gibson, 2004) is triggered by exothermic prograde accretion of extreme Schwarzschild (non-spinning) black holes on extreme (spinning) Kerr black holes (Peacock, 2000, p61).

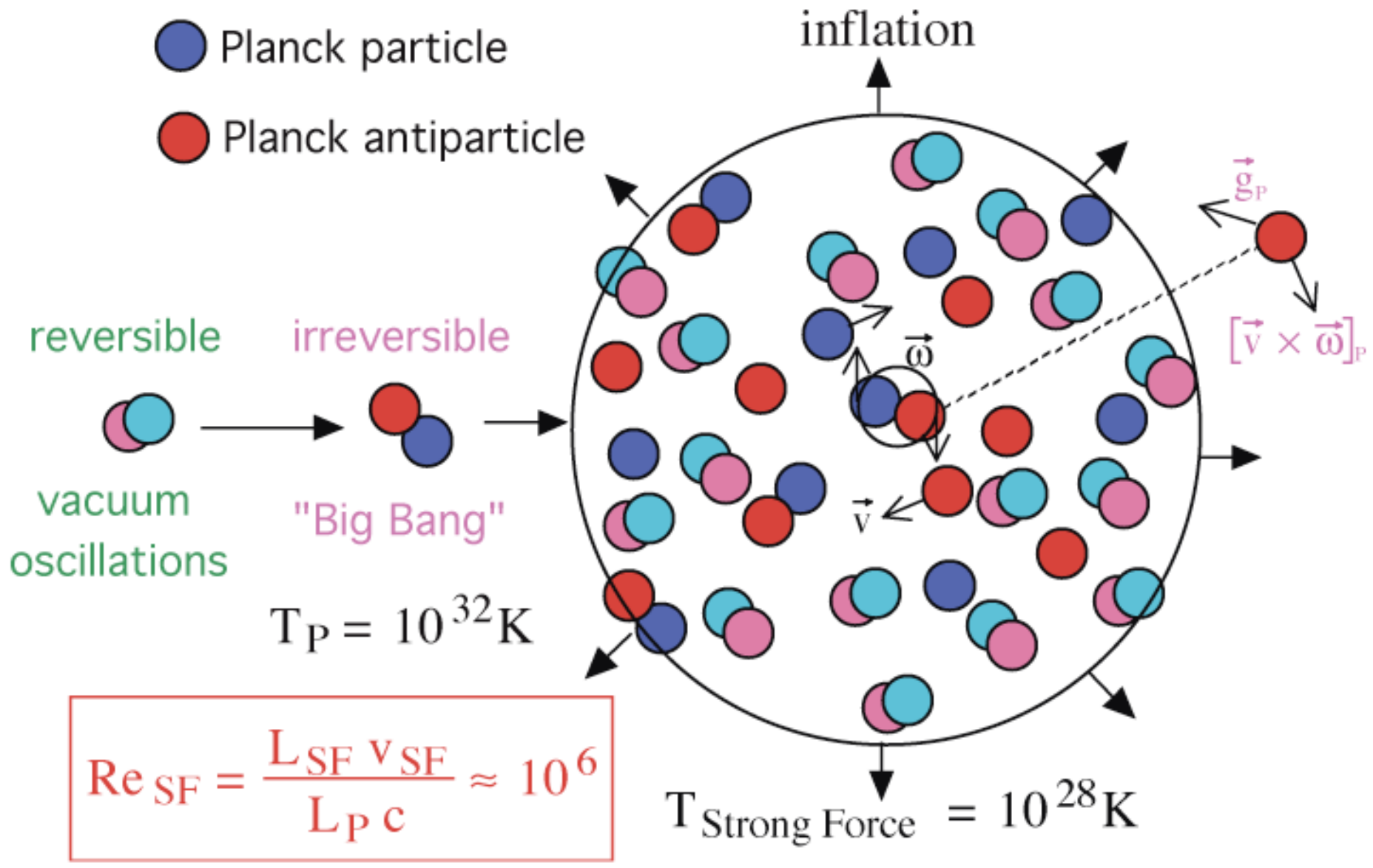

Figure 2. Physical process of big-bang-turbulence-combustion (Gibson, 2004).

In Fig. 2, vacuum oscillations produce a Planck particle-antiparticle pair by quantum tunneling. Quantized rotation states (extreme Kerr black holes, center) form. Inertial-vortex-forces (upper right) balance gravity, diffuse the spin, slow the annihilation, and homogenize the velocity and temperature fluctuations by turbulent eddy formation and mixing until the strong-force-freeze-out (GUT) temperature $T_{SF}$ is reached. Exponential inflation of space results from the Reynolds stress $\tau_P$, gluon-viscous-stress and false-vacuum-pressure (Guth, 1997). The turbulent temperature field fossilizes by expansion beyond the scale of causal connection *ct*. The particles in Fig. 2 are not to scale. Kerr particles are smaller than the Planck particles and prograde orbit daughters are smaller than their parents, reflecting the release of gravitational potential energy to fuel big bang turbulent combustion.

In our model, QGD momentum transport (Fig. 2) is only by short range Planck particle interactions with small kinematic viscosities $\nu_P \approx cL_P$ and large Reynolds numbers $Re = vL/cL_P$ at scale $L$ with velocity $v \approx (2kT)^{1/2}$. The big-bang-turbulence cascade terminates when temperatures decrease to the (GUT) strong-force-freeze-out temperature $T_{SF} = 10^{28}$ K at time $10^{-35}$ seconds. Quarks and gluons form with $L = L_{SF} = 10^8 L_P$, $v_{SF} = 10^{-2} c$ giving a maximum Reynolds number of order $10^6$ as shown in Fig. 2.

Negative pressures produce space-time-energy from Einstein's general relativity theory equations. These connect space-time-geometry to the stress-energy tensor (Peacock, 2000, p20)

$$G^{\mu\nu} = R^{\mu\nu} - g^{\mu\nu}R = -\frac{8\pi G}{c^4}T^{\mu\nu} = -\frac{8\pi}{F_P}T^{\mu\nu} \quad (3)$$

where the Einstein tensor on the left of (3) is shown in terms of the Ricci tensor $R^{\mu\nu}$, the metric tensor $g^{\mu\nu}$ and the curvature scalar $R$ (Weinberg, 1972). Big-bang-turbulence theory (Gibson, 2004) requires that the usual assumptions of ideal, collisionless, adiabatic, linear fluids be abandoned so that the stress-energy tensor $T^{\mu\nu}$ on the right of (3) can include viscous and turbulent Reynolds stresses and turbulent entropy production. Large distortions of space-time described by Kerr and the Schwarzschild metric tensors near the interacting extreme black holes is a subject for future work.

After exponential inflation of space by a factor of order $10^{25}$ (Guth, 1997) the big bang fossil-temperature-turbulence fluctuations continue to stretch as the universe expands, from general relativity theory. Fossil Planck-scale temperature fluctuations reenter the horizon first because they have the smallest scales, but not until after the universe cools to the electoweak freeze-out temperature $3 \times 10^{15}$ K so that radiation (momentum transport by active neutrinos and photons with collision paths larger than electron separations but smaller than the horizon) can provide viscosities sufficient to damp out any turbulence (Gibson, 2000). The fluctuations trigger nucleosynthesis in patterns of hydrogen and helium density reflecting the extreme temperature sensitivity of this process (Peacock, 2000).

The anisotropy of the spinning turbulent instability mechanism may account for the excess of baryonic particles versus anti-particles observed in the universe, and a possible sterile neutrino population that seems to be the most likely non-baryonic dark matter candidate (Fuller et al., 2003). Because the geometry of the universe appears to be flat from observations and the “dark energy” hypothesis implausible and unnecessary, a massive non-baryonic dark matter component can be inferred since nucleosynthesis does not permit a sufficient baryonic component to balance the observed rate of expansion. This weakly collisional material that accounts for 97% of the universe mass manifests itself only by gravitational effects. For example, when supercluster mass baryonic fragments appear at $10^{12}$ s in the primordial plasma as viscous forces of the expansion match gravitational forces at the horizon scale, the non-baryonic dark matter immediately diffuses to fill the protosupercluster voids and decrease the gravitational driving force. This explains why the CMB temperature anisotropies are so small, with $\delta T/T$ values of order $10^{-5}$, even though gravitational plasma fragmentation is well advanced at the CMB time of $10^{13}$ s.

At this plasma-gas transition, atoms formed and photons decoupled from electrons, giving the cosmic microwave background (CMB) image of the universe that we observe today, Figure 3, redshifted a factor of 1100 by the expansion of the universe into the microwave bandwidth from the original white-hot visible wavelengths to the observed temperature 2.7 K. Small temperature anisotropies are found in the rest mass frame of the big bang by correcting for Galaxy velocity. The viscosity decreased by $10^{12}$ and the gas turned to $10^{24}$ kg fog particles in $10^{36}$ kg clumps that persist as the baryonic dark matter (Gibson, 1996).

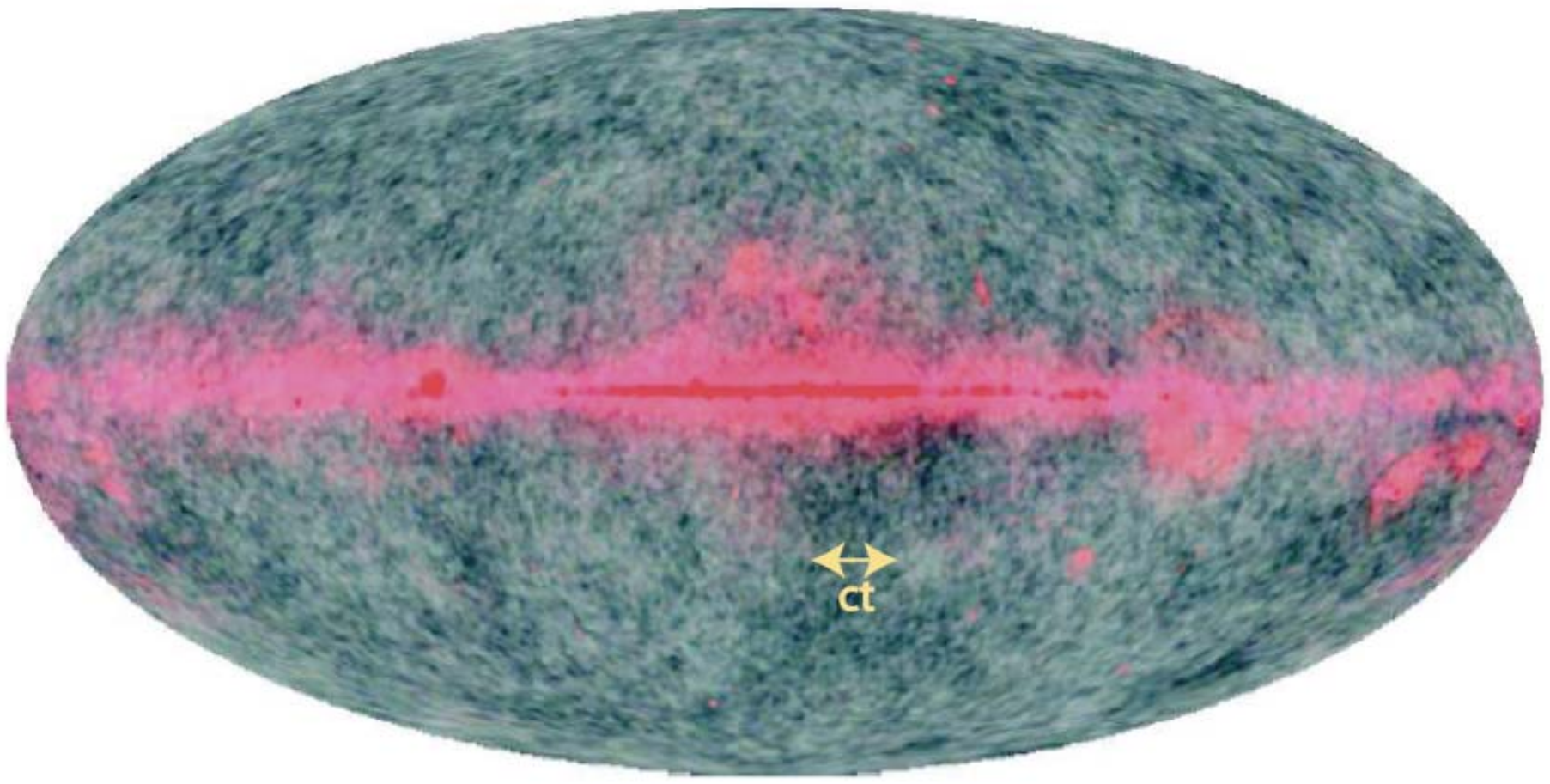

Figure 3. Sky map of CMB temperature anisotropies $\delta T/T \sim 10^{-5}$ from the WMAP satellite (Bennett et al., 2003). Except for the strong equatorial Milky Way Galaxy noise, all the fluctuations are chaotic, homogeneous and isotropic at scales larger that the horizon scale *ct* existing at the 300,000-year time of the CMB (as shown by the double arrow) consistent with a big-bang-turbulence-combustion origin.

## DISCUSSION OF CMB SPECTRA

A variety of high resolution maps and spectra of Cosmic-Microwave-Background (CMB) temperature fluctuations have been obtained from telescopes on earth and on balloons and spacecraft at altitudes above atmospheric interference, extending the 1989 Cosmic-Background-Experiment (COBE) space telescope observations to smaller scales and higher precision. Hu 2000 presents a collection of spectral measurements and CDM models. These are compared to our big-bang-turbulence-combustion predictions and more recent data in Figure 4 (Gibson, 2004). The COBE point on the left (missing from the Hu 2000 Figure) is interpreted as the fossil strong-force (GUT) horizon scale $L_{HSF}$ stretched and inflated by a factor of $10^{50}$. WMAP (red triangles) confirms this COBE datum. Rather than drooping at high wavenumbers as predicted by acoustic models and magnetic models, BBT predicts a spectral cut off at a stretched and inflated Planck scale ($L_{Planck}$ x $10^{50}$) four decades to the right as shown by the arrow. Radio-telescope-array results (CBI, ACBAR, BIMA) support this BBT prediction for $10^3 < k < 10^4$, (Readhead et al., 2004) with 98% confidence, suggesting the excess power detected may be from secondary Sunyaev-Zeldovich anisotropies in distant galaxy clusters (ruled out by HGD).

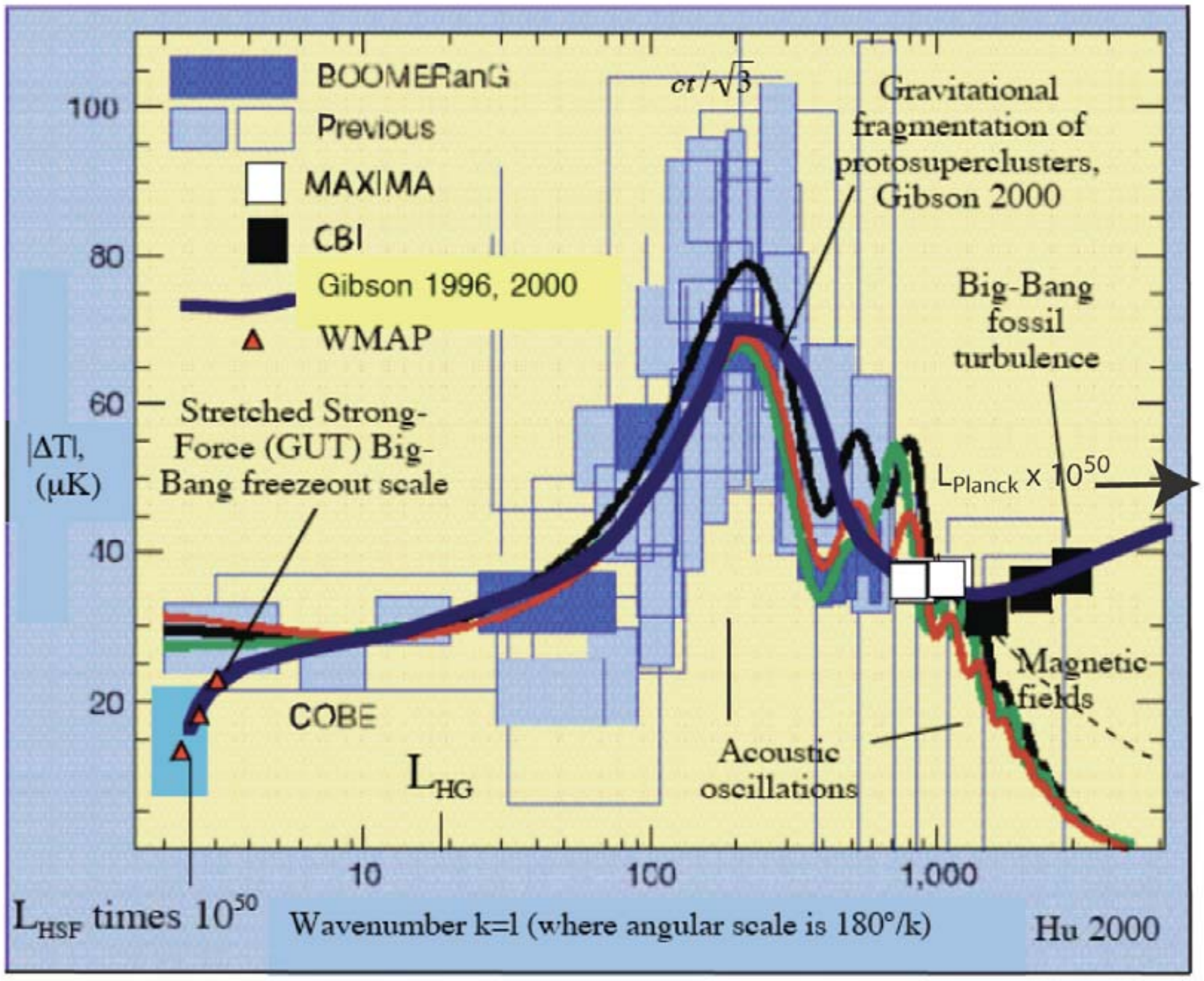

Figure 4. Measured cosmic-microwave-background (CMB) temperature gradient power spectra ($|\delta T|(k)=[k^2\phi(k)]^{1/2}$, μK) compared to three cold-dark-matter (CDM) acoustical models and the present predictions of early structure and big-bang fossil turbulence $\sim k^{1/6}$ (k = l).

The predicted GUT big-bang-turbulence (BBT) spectral cut-off confirmed by COBE and WMAP data on the left in Fig. 4 contradicts cold-dark-matter CDM models. COBE (pale), BOOMERanG (dark), MAXIMA (white), and CBI (black, Pierson et al., 2003) data points show a maximum at k = 220. The spectral peak occurs at a sonic wavelength $\lambda = ct/3^{1/2}$ for time $t = 10^{13}$ s, where $c$ is the speed of light. This has been interpreted by astrophysical authors as proof that plasma is collected by gravity into $\sim 10^{36}$ kg collisionless-CDM-clump gravitational potential wells where it oscillates acoustically. These "CDM halos" later hierarchically cluster to form galaxies, clusters and superclusters in the standard CDMHC structure formation model. Such CDM halos are diffusionally unstable (Gibson, 2000). Weakly collisional (CDM) particles would rapidly diffuse away from such clumps since the diffusivity for weakly collisional particles is large. A continuous powerful sound source is needed to produce a sonic peak with $\delta T/T \approx 10^{-4}$. Sonic oscillations in the primordial plasma are rapidly damped by photon-viscous forces due to Thomson scattering of photons with electrons that are strongly coupled to ions by electric forces. The photon viscosity of the plasma at the time of first structure formation $10^{12}$ s is estimated to be $4\times10^{26}$ $m^2$ $s^{-1}$ (Gibson, 2000), giving a Reynolds number ~$10^2$.

Instead, the spectral peak more likely reflects the first gravitational formation of structure in the plasma epoch by hydro-gravitational-dynamics (HGD) criteria (Gibson, 1996, 2000, 2004) that replace the Jeans 1902 acoustical-criterion by including viscosity, turbulence and diffusion. Then it is no coincidence that the wavenumber of the peak is at an acoustical scale. This is because structure forms by expansion of density minima to form voids, and because rarefaction waves of the expanding voids are limited to the sound speed. As shown in Figure 5, the universe expands following the big bang. Nucleosythesis of hydrogen, helium and electrons and a large mass of non-baryonic (neutrino-like) dark matter (NBDM) occur in BBT patterns. Gravitational structure forms when the viscous-gravitational scale matches the horizon scale. These plasma proto-supercluster voids fill with the NBDM material by diffusion, reducing the gravitational driving force of structure formation. Proto-cluster-voids and proto-galaxy-voids expand until the plasma cools to about 3000 K, so that gas and the CMB can appear. Gas is transparent to photons, which is why we can see the state of the universe back to 300,000 years by means of the CMB in Fig 3. From Fig. 5 the large scale structures of the universe are already in place at the time of the plasma-gas transition. Once structures form at $10^{12}$ seconds, the baryonic density $\rho_0 \sim 10^{-17}$ kg $m^{-3}$ existing then is preserved as a fossil, and appears as the density of the proto-globular-star-clusters (PGCs) and primordial-fog-particles (PFPs) that form in the primordial gas of proto-galaxies.

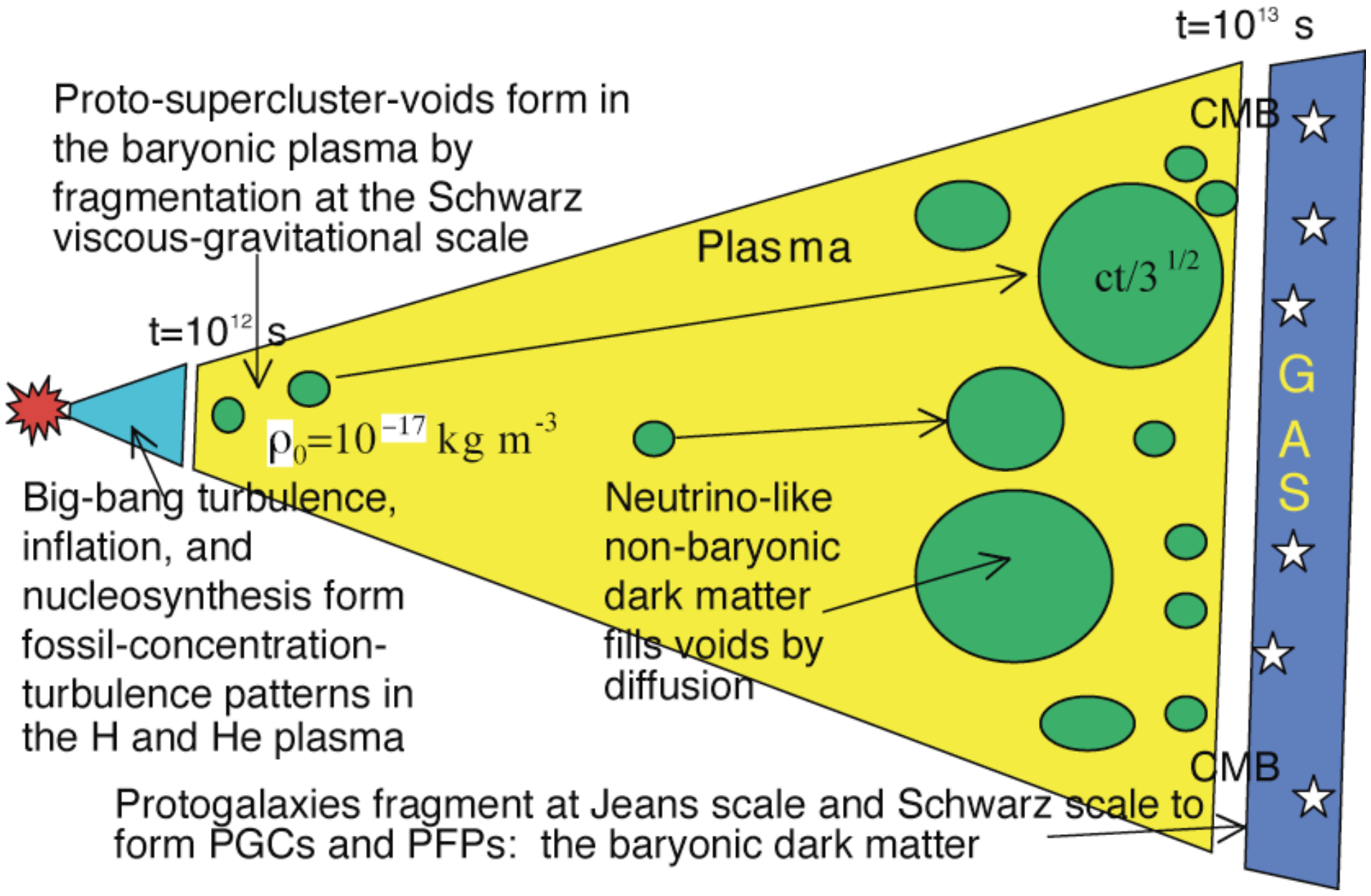

Figure 5. Hydro-gravitational structure formation (Gibson, 1996) during the plasma epoch between $10^{12}$ seconds (30,000 years) and the plasma-gas transition at $10^{13}$ seconds (300,000 years). The protosupercluster voids expand at sound speed $V_s = c/3^{1/2}$ to the scale $ct/3^{1/2}$ of the acoustic peak in Fig. 4. The non-baryonic dark matter diffuses to fill the voids and slow the gravitational fragmentation. After transition to gas, planets and stars form in proto-globular-star clusters within the protogalaxies in a free-fall time $(\rho_0 G)^{-1/2} \sim 10^{13}$ s (300,000 years) set by the fossilized density $\rho_0 \sim 10^{-17}$ kg m$^{-3}$ from the time of first structure.

Proto-galaxies are the smallest gravitational structures formed in the plasma epoch, Fig. 5. Buoyancy forces from self-gravity in the plasma structures damp turbulence and preserve the density and rate-of-strain existing in the plasma at $10^{12}$ seconds. Stars appear in a free fall time $10^{13}$ s by accretion of PFPs, first at the centers of PGCs near the centers of protogalaxies. There is no "dark age" period of 300 million years before the first star appears as required by the Jeans 1902 criterion. The extremely gentle, nonturbulent condition of the primordial gas permits the formation of small, ancient stars observed in the uniform population of $10^{36}$ kg spherically-symmetric globular-star-clusters found with density $\rho_0$ in all galaxies. Reionization of the universe by Population III superstars never happened. Hierarchical clustering of CDM halos to form galaxies never happened. Hierarchical clustering of galaxies to form galaxy clusters and superclusters never happened.

Significant distortion of space-time due to GR must be considered in the interpretation of CMB data as evidence of big bang turbulence, as shown in Figure 6.

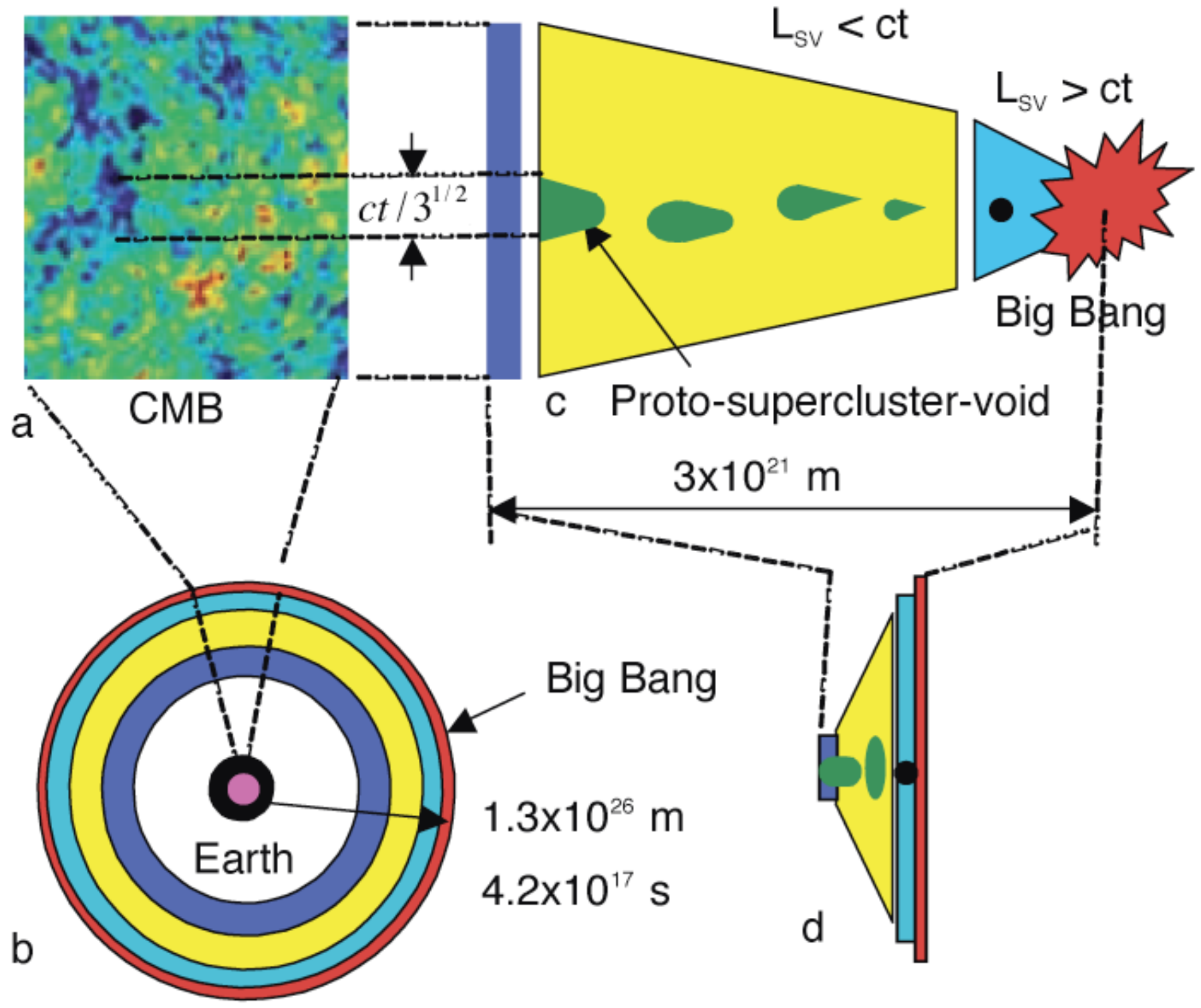

Figure 6. General relativity affects the interpretation of the CMB temperature anisotropies shown in the upper left (a) as observed from Earth (b). Hydro-gravitational structure formation regions of Fig. 5, (c), appear stretched into thin spherical shells (b, d). The $3\times10^{25}$ m diameter ($10^3$ Mpc) black sphere surrounding Earth in (b) contains about $10^4$ galaxy super-clusters.

A small patch of the WMAP CMB image is shown in Fig. 6a, as observed from Earth in Fig. 6b. Looking outward is looking back in time. The radius of the outer sphere is our horizon scale $ct=4\times10^{26}$ m for an assumed age of the universe 13.3 billion years, or $4.2\times10^{17}$ s, as shown. The dark blue regions of the CMB in (a) represent proto-supercluster-void regions of density deficiency, seeded by BBT fossils in the nucleosynthesis epoch (black dots in c, d), that have expanded at sub-sonic acoustic speeds < $c/3^{1/2}$ to form the growing voids of (c) filled with "neutrinos" by diffusion. From GR, light from these regions is blue-shifted by gravity compared to the proto-supercluster regions in red (a) that are red-shifted. The black sphere around Earth in Fig. 6b contains galaxies up to 12 billion years old with red shift $z < 0.1$. The gas (blue), plasma (yellow), energy-dominated (turquoise), and big-bang (red) shells have been thickened by factors > $10^4$ to make them visible. Super-galaxy-cluster-void bubbles in Fig. 6c are stretched by GR but preserve patterns reflecting fossil-turbulence-density-minimum seeds from nucleosynthesis, black dots in Fig. 6c and 6d. As the voids grow they are filled by strongly-diffusive non-baryonic dark matter material that reduces the gravitational driving force. This is proposed to be a massive soup of relic neutrino flavors from the viscous, energy dominated, nucleosynthesis epoch (Fuller et al., 2003). Gravitational structure formation begins in the plasma epoch when $L_{SV}<ct$, as shown. It is prevented by viscous forces in the energy epoch where $L_{SV}>ct$.

The Jeans 1902 acoustic gravitational instability criterion is the basis of the cold-dark-matter (CDM) hypothesis of hierarchical galaxy structure formation. Jeans' criterion is derived by a linear perturbation stability analysis of the inviscid Euler equations with gravity and neglecting diffusion, which reduces the stability problem to one of gravitational acoustics. By the Jeans criterion, sound waves are unstable in a gas of density $\rho$ if the time $L/V_S$ required to propagate a wavelength $L$ at sound speed $V_S$ is greater than the time required for gravitational free fall $(\rho G)^{-1/2}$. Density fluctuations smaller than the Jeans scale $L_J = V_S(\rho G)^{-1/2}$ are assumed to be gravitationally stable in standard cosmological models (Weinberg, 1972; Silk, 1989; Kolb and Turner, 1990; Peebles, 1993; Padmanabhan, 1993 and Rees, 2000), where $G$ is Newton's gravitational constant (Fig. 1). However, analysis using viscous, turbulent, and other forces as well as diffusion gives gravitational structure formation at Schwarz scales (i.e.: $L_{SV}$, $L_{ST}$, $L_{SD}$) that may be smaller or larger than $L_J$ (Gibson, 1996).

By hydro-gravitational-dynamics (HGD), proto-galaxies fragment (for heat transfer reasons) to form Jeans-mass $10^{36}$ kg proto-globular-star-clusters (PGCs) of planetary mass fog particles (PFPs) at $L_{SV}$ scales, as observed by quasar microlensing (Schild, 1996) and in space telescope images of planetary nebulae (Gibson and Schild, 2003). Viscous and turbulence forces determine gravitational structure formation when either the viscous Schwarz scale $L_{SV}$ or the turbulent Schwarz scale $L_{ST}$ become smaller than $L_H$, where $L_{SV} = (\gamma\nu/\rho G)^{1/2}$, $L_{ST} = \varepsilon^{1/2}/(\rho G)^{3/4}$, $\gamma$ is the rate of strain of the gas with density $\rho$ and kinematic viscosity $\nu$, and $\varepsilon$ is the viscous dissipation rate (Gibson, 1996). All the structures of the plasma epoch in Fig. 5 monotonically expand in the gas epoch at rates, determined by friction, smaller than the Hubble "constant" strain-rate of space $\gamma \approx 1/t$. Stars form by a binary gravitational-accretion-cascade of about 3% of the PFP planets. Most of the PFP planets have frozen to form the baryonic dark matter. The non-baryonic dark matter has diffused to form the outer halos of galaxy clusters and isolated galaxies at diffusive Schwarz scales $L_{SD} = (D^2/\rho G)^{1/4}$ more than $10^{22}$ m, >10 times larger than the baryonic dark matter halo scale of galaxies, with <10 times smaller density but >10 times larger mass.

## STRUCTURE FUNCTION AND ESS COEFFICIENTS OF THE CMB

Power spectra of the CMB temperature anisotropies such as Fig. 4 utilize only a small fraction of the information contained in the data. To investigate the possibility that the fluctuations have a turbulence origin, more sophisticated statistical parameters are useful such as structure function coefficients and extended-self-similarity (ESS) coefficients. Structure function coefficients $\zeta_p$ of cosmic microwave background temperature anisotropy data (www.hep.upenn.edu/ ~xuyz/qmask.html, Xu et al. 2001) have been computed (Bershadskii and Sreenivasan, 2002), where $\zeta_p$ are power law coefficients for $p^{th}$ order structure functions $|\Delta T|_r^p \approx r^{\xi_p}$ for fluctuations $\Delta T(r)$ of hydrophysical fields like temperature T with sampling point separation $r$.

Results are shown in Figure 7 compared to $\zeta_p$ values from the low Reynolds number magneto-hydro-dynamic (MHD) flow of the solar wind (Benzi et al., 1996, Table 1). The close agreement between the $\zeta_p$ coefficients for the CMB and for a high Reynolds number turbulent flow can hardly be a coincidence. Because the $\delta T/T$ fluctuations are so small, it is not possible that the plasma was strongly turbulent at the CMB time of sampling $10^{13}$ s. Therefore, the turbulence signature must either be from the plasma epoch or from the big bang itself. Estimates of the Reynolds numbers from large photon viscosities of the plasma epoch are too small (Gibson, 2000) to permit the strong turbulence suggested by Fig. 7.

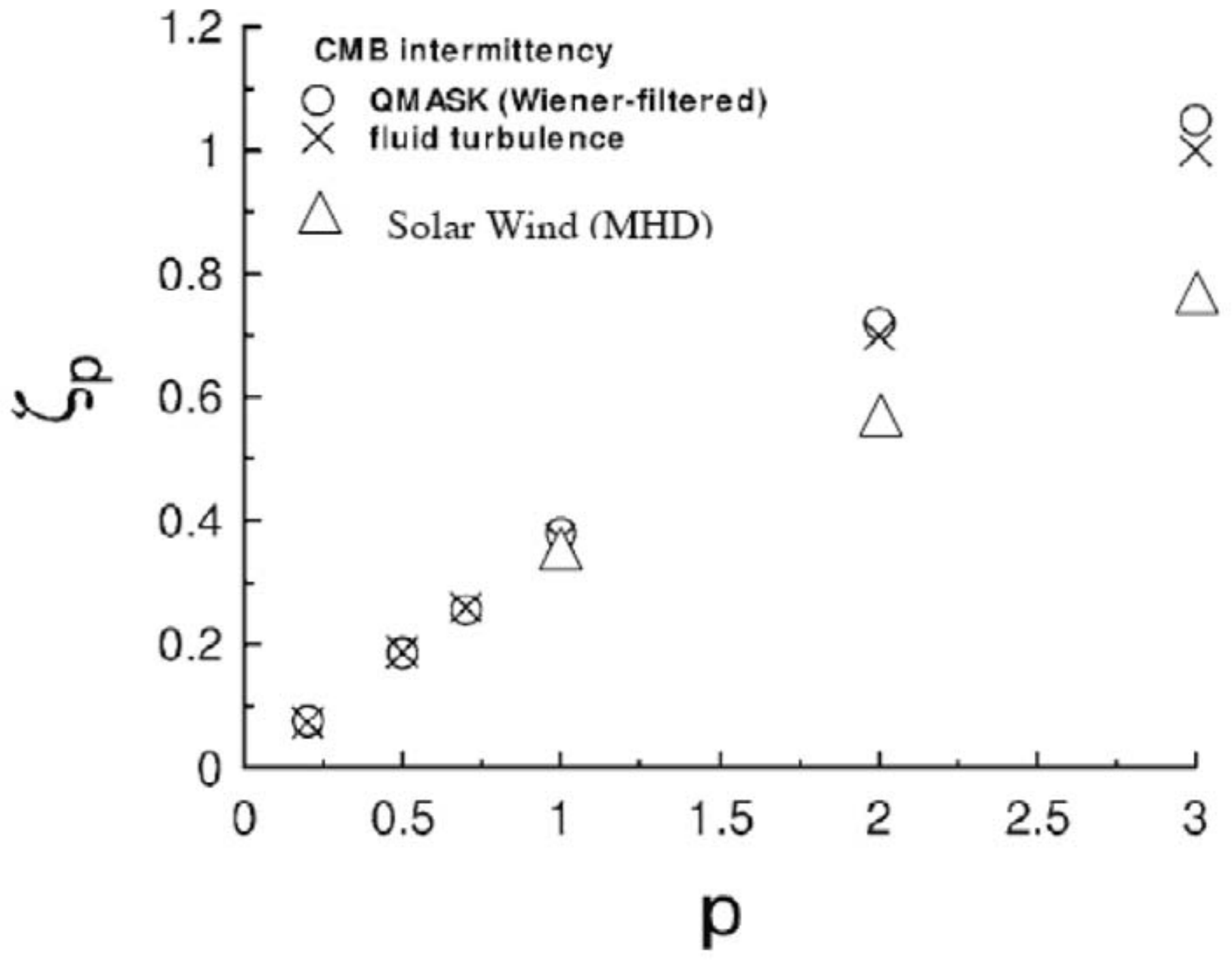

Figure 7. Structure function coefficients $\zeta_P$ for CMB temperature anisotropies (circles, Bershadskii and Sreenivasan, 2002) compared to those for high Reynolds number fluid turbulence (×) and the low Reynolds number MHD solar wind (Benzi et al., 1996).

Perhaps the best-known velocity-structure-function-relation for high Reynolds number turbulence velocity differences was predicted by Kolmogorov in 1941. Using his second universal similarity hypothesis for p = 3 Kolmogorov found $\zeta_p$ = 1 by dimensional analysis. For p = 2 one finds $\zeta_p$ = 2/3 by the same method, corresponding to the -5/3 inertial subrange of universal velocity and turbulent mixing spectra. These values are shown in Fig. 7 for high Reynolds number turbulence and the CMB temperature anisotropies, but not for the low Reynolds number MHD fluctuations of the solar wind (Benzi et al., 1996).

Ground based and balloon observations of Fig. 7 were compared (Bershadskii and Sreenivasan, 2002) to be sure the turbulence signature was not from the atmosphere of the earth. Further tests of the CMB temperature anisotropies for turbulence signatures have been made with even higher accuracy and spatial resolution of the Wilkenson Microwave Anisotropy Probe (WMAP) using the Extended-Self-Similarity (ESS) method (Benzi et al., 1996), as shown in Figure 8 (Bershadskii and Sreenivasan 2003). ESS coefficients are formed by ratios of structure functions of various orders. The expression used in Fig. 8 is

$$\left\langle |\Delta T_r|^p \right\rangle \sim \left\langle |\Delta T_r|^3 \right\rangle^{\varsigma_P} \quad (4)$$

from Kolmogorov's second law and dimensional analysis. For a Gaussian process, the exponent $\varsigma_P = p/3$. The departure from the Gaussian curve for p>4 is identical for both the WMAP data and for turbulence data, confirming the previous evidence from Fig. 7 and Fig. 4 that the CMB temperature anisotropies preserve strong evidence of a primordial turbulence origin.

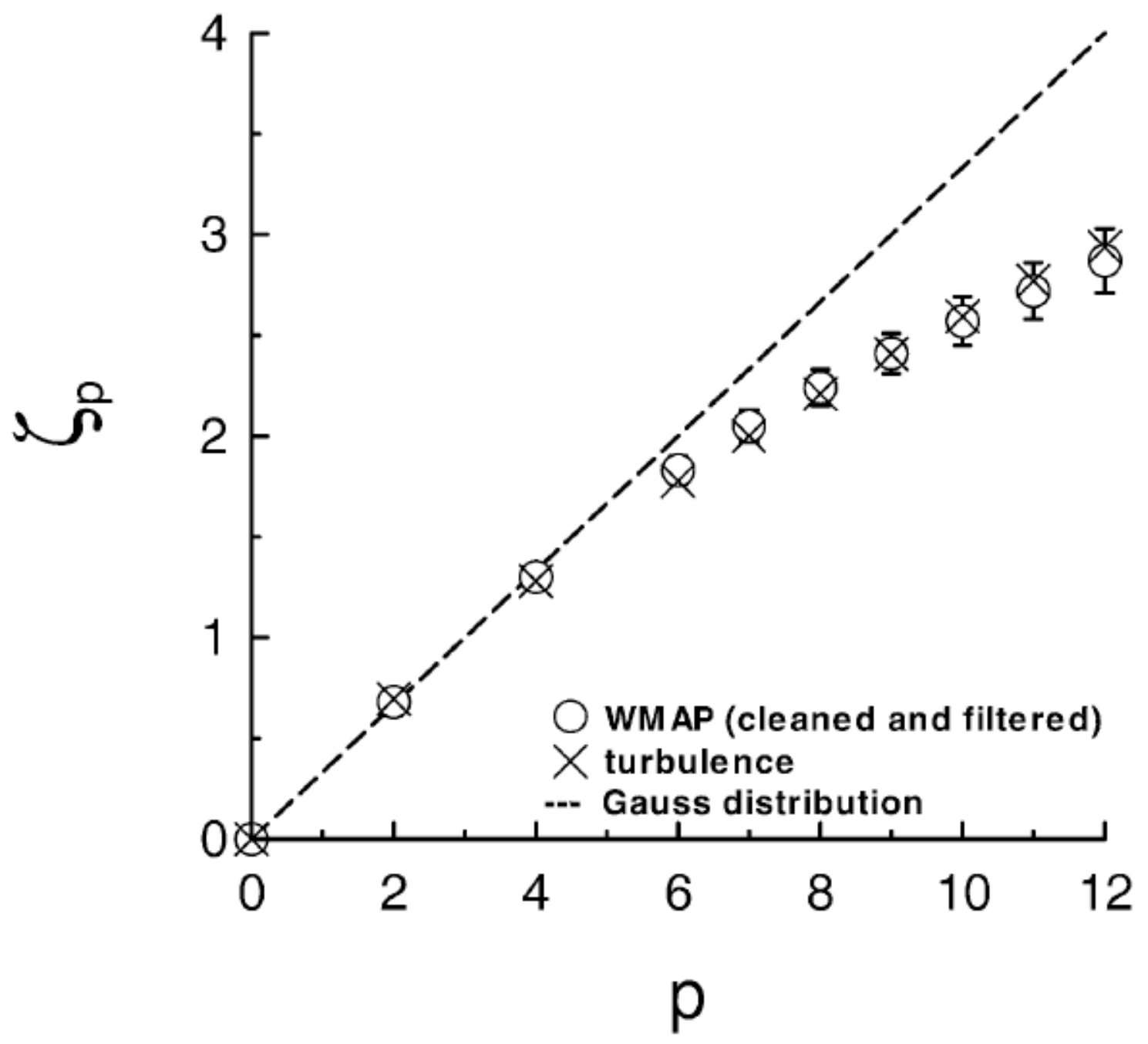

Figure 8. Extended Self Similarity (ESS) coefficients computed from WMAP temperature anisotropies compared to those of turbulence (Bershadskii and Sreenivasan, 2003).

The evidence in Figs. 7 and 8 shows the fingerprints, if not DNA evidence, of high Reynolds number turbulence (HRT) in the cosmic-microwave-background data. No epoch preceding the CMB plasma-gas transition permits HRT other than the big bang.

## CONCLUSIONS

A chaotic, quantum-gravity, big-bang-turbulence-combustion model at Planck scales satisfies our narrow definition of turbulence and the small-to-large direction of the turbulent cascade. Inertial-vortex forces are identified with an efficient, but entropy producing, Hawking radiation of Planck-particle mass-energy and angular momentum that results when extreme-Schwarzschild-black-holes achieve minimum-radius prograde-orbits about spinning extreme-Kerr-black-holes. The Planck inertial-vortex-force matches the Planck scale gravitational force and produces the vorticity, Planck-particle gas, big-bang-turbulence-combustion and entropy-irreversibility required to form the universe. At Planck temperatures $10^{32}$ K, only Planck particles, Planck antiparticles, and Planck-Kerr particles can exist. These interact at a Planck-length-scale $L_P = 10^{-35}$ m that is less than the expanding universe horizon-scale $L_H = ct$. Only Planck particle-particle viscosities $\nu_P = cL_P$ can transmit momentum, thus giving large Reynolds numbers $L_{SF}v_{SF}/L_P c$ up to $10^6$ in the big bang turbulence before the universe cools to the GUT strong-force-freeze-out temperature $10^{28}$ K. Quarks and gluons can then appear with $10^6$ larger gluon-viscosity $L_{SF}v_{SF}$ that damps the big-bang turbulence, with collision length $L_{SF} = 10^8 L_P$ and velocity $v_{SF} = 10^{-2}c$.

Negative gluon-viscous-stresses, negative pressures of the turbulent-Reynolds-stress, and possibly a Guth negative false-vacuum pressure combine to produce an exponential inflation of space from Planck to atomic dimensions according to Einstein's general relativity equations. Evidence of big bang turbulence is provided by the CMB in several forms. The spectral peak in Fig. 4 supports the HGD prediction that the first gravitational structures should be proto-supercluster-voids expanding at acoustic velocities $c/3^{1/2}$ from density minima starting at about $10^{12}$ s to $ct/3^{1/2}$ with $t=10^{13}$ s, as shown in Figs. 5 and 6. High wavenumber Cosmic-Background-Imager (CBI) spectral data show the spectrum is not CDM and not magnetic. Structure function and ESS coefficients of Figs. 7 and 8 show statistical parameters in very good agreement with those of high Reynolds number turbulence. This supports a big-bang-turbulence-combustion scenario, since turbulence in the photon-viscous plasma-epoch at the time of first gravitational structure formation (Gibson, 1996, 2000) at 30,000 years ($10^{12}$ s) has low Reynolds numbers $\sim 10^2$. Formation of the first plasma structures by fragmentation at proto-galaxy-supercluster mass scales $10^{46}$ kg causes buoyancy forces due to self-gravity. Buoyant damping can explain the lack of any turbulence at plasma-gas transition ($10^{13}$ s) indicated by the small $\delta T/T \sim 10^{-5}$ K CMB temperature anisotropy levels. Viscous damping would require $\nu \sim 10^{30}$ m$^2$ s$^{-1}$, which is much larger than the photon viscosity $10^{25}$ m$^2$ s$^{-1}$ expected for the plasma.

The WMAP small k cutoff in Fig. 4 suggests that remnants of big-bang-turbulence at the GUT strong-force-freeze-out scale $L_{SF} \sim 10^{-27}$ m were stretched by a factor of about $10^{50}$ to scales ten times larger than the gas-horizon wavelength $L_{HG}$ existing at $10^{13}$ s. Big-bang-turbulence occurred before cosmological inflation, which explains the random, homogeneous, isotropic, galaxy density fluctuations observed to be independent of direction on the sky but outside each other's horizon range of causal connection *ct*.

The CMB dominant spectral peak at $3 \times 10^{21}$ m wavelength has expanded by a factor of $10^3$ corresponding to the CMB redshift to the present observed size of galaxy-supercluster-voids. From hydro-gravitational-dynamics, the CMB peak is due to plasma gravitational-structure-formation by fragmentation, not sonic oscillations in CDM halos. Sound speed is the maximum limit for the rarefaction waves of gravitational void formation in the plasma from the second law of thermodynamics. Galaxy-supercluster-voids and the voids between galaxy-clusters and galaxies have expanded from gravity forces and with all other space in the universe according to general relativity theory. Galaxy-clusters and galaxies have expanded less rapidly because their expansion is inhibited by gravity and by gas friction.

Gas friction is provided by evaporation of the baryonic dark matter, which is a large population of rogue Earth-mass frozen-gas planets (PFPs) in dark proto-globular-star-cluster (PGC) clumps and disrupted into the Galaxy core and disk (Gibson, 1996). Quasar-microlensing by a lens-galaxy at planet-mass twinkling frequencies supports this interpretation (Schild, 1996). Stars are formed from PFP planets by a binary gravitational accretion cascade to ever increasing planet mass. Thus the interstellar medium (ISM) is thin gas evaporated from ~30 million PFP rogue planets surrounding the average star and dominating the $10^{-17}$ kg m$^{-3}$ ISM density. The present average mass density for a flat universe is $\rho_c = 10^{-26}$ kg m$^{-3}$, compared to $\rho \approx 10^{-21}$ kg m$^{-3}$ for a galaxy, $10^{-22}$ kg m$^{-3}$ for a galaxy-cluster and $10^{-23}$ kg m$^{-3}$ for a supercluster, all starting from an initial density of $\rho_0 = 10^{-17}$ kg m$^{-3}$ as shown in Fig. 5. The density of globular-star-clusters and PGCs match this initial density $\rho_0$. We conclude the density $\rho_0 = 10^{-17}$ kg m$^{-3}$ is a fossil of the first gravitational structure formation in the primordial plasma with this density beginning at $10^{12}$ seconds (30,000 years) when the horizon mass was that of a supercluster.

The CDM hierarchical clustering (CDMHC) model predicts that galaxy superclusters are the last stage of gravitational structure formation after the big bang (rather than the first stage as predicted by HGD). Its basis is the obsolete and incorrect fluid mechanical theory of Jeans 1902, and the unwarranted assumption that "collisionless fluid mechanics" and the collisionless Boltzmann equation apply to the non-baryonic dark matter (or any other real fluid). All recent observations show conflicting evidence of early stars, galaxies and galaxy clusters. The CDMHC model and the CDM concept of clumped diffusive fluid should be abandoned.

## ACKNOWLEDGEMENT

It is a pleasure to acknowledge, on the occasion of his $70^{th}$ birthday celebration, many valuable conversations with Forman Williams about turbulence, turbulent mixing, and combustion, during our many years together at UCSD, that have contributed to this paper.